\documentclass{aa}
\usepackage{graphicx}
\begin{document}

%
\title{{\em XMM-Newton} observation of the Lockman Hole
        \thanks{Based on observations obtained with {\em XMM-Newton}, an ESA 
        science 
           mission with instruments and contributions directly funded by 
           ESA Member States and the USA (NASA)}}

\subtitle{I. The X-ray Data}

   \author{G. Hasinger\inst{1}
           \and B. Altieri\inst{2} 
           \and M. Arnaud\inst{3} 
           \and X. Barcons\inst{4}
           \and J. Bergeron\inst{5}
           \and H. Brunner\inst{1} 
           \and M. Dadina\inst{6} 
           \and K. Dennerl\inst{7}
           \and P. Ferrando\inst{3} 
           \and A. Finoguenov\inst{7}  
           \and R. E. Griffiths\inst{8}
           \and Y. Hashimoto\inst{1}  
           \and F. A. Jansen\inst{9}
           \and D. H. Lumb\inst{9}
           \and K. O. Mason\inst{10} 
           \and S. Mateos\inst{4} 
           \and R. G. McMahon\inst{11} 
           \and T. Miyaji\inst{8}  
           \and F. Paerels\inst{12} 
           \and M. J. Page\inst{10}
           \and A. F. Ptak\inst{8}  
           \and T. P. Sasseen\inst{13}  
           \and N. Schartel\inst{2} 
           \and G. P. Szokoly\inst{1} 
           \and J. Tr\"umper\inst{7}
           \and M. Turner\inst{14}
           \and R. S. Warwick\inst{14}
           \and M. G. Watson\inst{14}
}

   \offprints{G. Hasinger, \email{ghasinger@aip.de}}

   \institute{Astrophysikalisches Institut Potsdam (AIP), 
              An der Sternwarte 16, D-14482 Potsdam, Germany,
         \and
              XMM-Newton Operations Centre, European Space Agency, Vilspa,
              Apartado 50727, E-28080 Madrid, Spain
         \and 
             CEA/DSM/DAPNIA/Service d'Astrophysique, CEA/Saclay,
             F-91191 Gif sur Yvette Cedex, France 
         \and
             Instituto de Fisica (Consejo Superior de
             Investigaciones Cientificas-Universidad de Cantabria), E-39005
             Santander, Spain
         \and
             European Southern Observatory, Karl-Schwarzschild-Stra{\ss}e 2,
             D-85748 Garching, Germany
         \and
             TeSRE-CNR, Via Gobetti 101, I-40129 Bologna, Italy 
         \and
             Max-Planck-Institut f\"ur extraterrestrische Physik,
             D-85740 Garching, Germany
         \and
             Department of Physics, Carnegie Mellon University, 5000 Forbes
             Ave., Pittsburgh, PA 15213, USA 
         \and
             Space Science Dept., European Space Agency, ESTEC, Postbus 299,
             2200 AG Noordwijk, Netherlands
         \and
             MSSL, University College London, Holmbury St Mary, Dorking,
             Surrey RH5 6NT, United Kingdom
         \and
             Institute of Astronomy, Madingley Road, Cambridge CB3 0HA,
             United Kingdom
         \and 
             Columbia Astrophysics Laboratory, Columbia University,
             538 West 120th Street, New York, NY 10027, USA
         \and
             Department of Physics, University of California at Santa Barbara,
             Santa Barbara, CA 93110, USA
         \and
             X-ray Astronomy Group, Department of Physics and Astronomy,
             Leicester University, Leicester LE1 7RH, United Kingdom\\
}

   \date{Received 30 September 2000 / accepted 24 October 2000}

   \titlerunning{{\em XMM} Observation of Lockman Hole}

   \authorrunning{Hasinger et al.}

\abstract{
We report on the first deep X-ray survey with the {\em XMM-Newton} observatory 
during the performance verification phase. The field of the Lockman Hole, one 
of the best studied sky areas over a very
wide range of wavelengths, has been observed.    
A total of $\sim$ 100 ksec good exposure time has been 
accumulated. Combining the images of the {\em European Photon Imaging
Camera} (EPIC) detectors we reach a flux limit
of 0.31, 1.4 and $2.4 \times 
10^{-15}~{\rm erg}~{\rm cm}^{-2}~{\rm s}^{-1}$, respectively in the
0.5-2, 2-10, and 5-10 keV band. Within an off-axis angle of 10 arcmin
we detect 148, 112 and 61 sources, respectively. The log(N)-log(S) relation
in the three bands is compared with previous results. In
particular in the 5-10 keV band these observations present the deepest X-ray
survey ever, about a factor 20 more sensitive than the previous 
{\em BeppoSAX} observations.
Using X-ray spectral diagnostics 
and the set of previously known, spectroscopically identified {\em ROSAT}
 sources
in the field, the new sources can be classified.    
{\em XMM-Newton} detects a significant number ($\sim$ 40\%) of X-ray 
sources with hard, probably intrinsically absorbed X-ray spectra, 
confirming a prediction of the population synthesis models for the X-ray 
background. 
}
\maketitle

%

\section{Introduction}
\label{sec:intr}

Deep X-ray surveys indicate that the cosmic X-ray background (XRB) is
largely due to accretion onto supermassive black holes, integrated over
cosmic time. In the soft (0.5-2 keV) band 80-90\% of the XRB flux 
has been resolved using {\em ROSAT} and recent {\em Chandra} surveys 
(Hasinger et al. \cite{hasi98a}, Mushotzky et 
al. \cite{mushotzky00}, Giacconi et al. \cite{giacconi00}). In the 
harder (2-10 keV) band 25-30\% of the background have been resolved 
in {\em ASCA} and {\em BeppoSAX} surveys (Ueda et al. \cite{ueda98}, 
Cagnoni et al.  \cite{cagnoni98}, Giommi et al. \cite{giommi00}), and 
more than 60\%, when the recent {\em Chandra} surveys are included.
Surveys in the very hard (5-10 keV) band have been pioneered using
{\em BeppoSAX} and resolve about 30\% of the XRB (Fiore et al.  
\cite{fiore99}). 
Those X-ray surveys with a high degree of completeness in optical
spectroscopy find predominantly Active Galactic Nuclei (AGN) as counterparts
of the faint X-ray source population 
(Bower et al. \cite{bower96}, Schmidt et al. \cite{schmidt98},
Zamorani et al. \cite{zamorani99}, 
Akiyama et al. \cite{akiyama00}), mainly X-ray and optically unobscured
AGN (type-1 Seyferts and QSOs) but also a smaller fraction of 
obscured AGN (type-2 Seyferts). Spectroscopic identifications of the 
{\em BeppoSAX} 
and {\em Chandra} surveys are still far from complete, however a mixture 
of obscured and unobscured AGN seems to be the dominant population 
in these samples, too (Fiore et al. \cite{fiore00}, Barger et al.
\cite{barger00}, Giacconi et al. \cite{giacconi00}). The most recent 
AGN X-ray luminosity function, derived from the {\em ROSAT} surveys, shows
evidence for luminosity-dependent density evolution and indicates 
a constant QSO space density at redshifts $2 < z < 4$ (Hasinger \cite{hasi98b},
Miyaji et al. 
\cite{miyaji00}) unlike optical QSO luminosity functions (Schmidt et al.
\cite{schmidt95}, Fan et al. \cite{fan00}). 

The X-ray observations are consistent with  
population synthesis models based on unified AGN schemes (Setti \& Woltjer 
\cite{setti89}, Madau et al. \cite{madau94}, Comastri
et al. \cite{comastri95}, Gilli et al. \cite{gilli99}), which explain 
the hard spectrum of the X-ray background by a mixture of absorbed and
unabsorbed AGN, folded with the corresponding luminosity function and its
cosmological evolution. According to these 
models most AGN spectra are heavily absorbed and about 80\% of the light 
produced by accretion will be absorbed by gas and dust   
(Fabian et al. \cite{fabian98}). However, these models are 
not unique and contain a number of hidden parameters, so
that their predictive power remains limited (e.g. Hasinger 
\cite{hasi00}). In particular they require a substantial contribution
of high-luminosity obscured X-ray sources (type-2 QSOs), which so far 
have not been detected in sufficient quantities (see the discussion in
Halpern et al. \cite{halpern99}). 
The large throughput and the unprecedented hard X-ray sensitivity of the 
telescopes aboard the recently launched {\em XMM-Newton} observatory
(hereafter {\em XMM}; Jansen et al. \cite{jansen01}) will ultimately yield 
spectra of the faint X-ray sources and constrain the evolution of their
physical properties, in particular the X-ray absorption.

Here we present results of the first deep survey taken with
{\em XMM} in the {\em Lockman Hole}, one of the best studied sky 
areas at all wavelengths. This paper concentrates on the X-ray 
data analysis. We show combined images from the 
EPIC pn-CCD (Str\"uder et al. \cite{strueder01}) and MOS CCD 
cameras (Turner et al. \cite{turner01}) and the derived source counts
in different energy bands. With the help of X-ray colour-colour diagrams
and the previously identified sources in this field
we show that it is possible to obtain a coarse source classification
based on {\em XMM} data alone.


\section{X-ray observations}
\label{sec:obs}

The Lockman Hole field, centered on the sky position RA 10:52:43, DEC +57:28:48 
(2000) was observed with {\em XMM} in five separate revolutions (70, 71, 73, 74 and 
81) during the period April 27-May 19, 2000 for a total exposure time of 190 
ksec. The pointing direction was changed slightly (by about 10$\arcsec$ in RA and DEC 
in order to bridge
the gaps between the CCD detectors between exposures) and the roll 
angle varied in the range 48.66 to 54.25 degrees between the different 
revolutions. The EPIC cameras were operated in the standard full-frame mode. 
The thin filter was used for the PN camera, while the thin and the thick filter
were alternated for the MOS1 and MOS2 cameras in order to obtain diagnostics 
about the soft proton particle background. Tab. \ref{tab:obs} gives a summary 
of the observations.

The PN and MOS data were preprocessed by the {\em XMM Survey Scientist Consortium}
(SSC; Watson et al. \cite{watson01}) using the {\em XMM} Standard Analysis System
(SAS) 
routines. The attitude and deadtime information was not available for the 
datasets but can be regarded stable enough to be assumed constant for each 
revolution. The preprocessed FITS events files were analysed using FTOOLS 
routines.

A substantial fraction of the observations was affected by high and flaring 
background fluxes with count rates up to several hundred per second, compared
to a quiet count rate of several counts per second per detector.  
The data were screened for low background intervals, rejecting times with a 
0.5-10 keV count rate higher than 8 cts/s for the PN and 3 cts/s for each of 
the two MOS cameras. The remaining good time intervals added up to 
about 100 ksec (see Tab. \ref{tab:obs}). The actual exposure time for the
three different detectors, pn-CCD, MOS1 and MOS2 were slightly different
due to the varying start and end times of individual observations. 
A number of hot pixels and hot columns were removed interactively from the 
events lists by identifying them in images accumulated in detector coordinates 
and then spatially filtering them out of the datasets. 
An Al-K$_\alpha$ line at 1.5 keV is present in both detector types.
The PN background spectrum
shows in addition a 
strong Cu-K$_\alpha$ line at 8.1 keV, which is not present in the MOS 
background. PN photons in the energy range 7.9-8.3 keV have therefore 
been neglected in the further analysis.

\begin{table}   
\caption[]{Observing Log}
\begin{center}
\begin{tabular}{lrrrrll}
\hline
\noalign{\smallskip}
Date & Start & End  & Dur. & Exp. & Filter & Filter \\
2000 & [UT]  & [UT] &  [sec]   &   [sec]  &  MOS1  &  MOS2  \\
\noalign{\smallskip}
\hline
\noalign{\smallskip}
 27.04 & 02:45 & 22:08 & 69761 & 31216 & thin  & thick \\
 29.04 & 02:37 & 22:00 & 69761 & 28608 & thick & thin  \\
 02.05 & 17:50 & 00:21 & 23512 & 12944 & thin  & thick \\
 05.05 & 08:48 & 21:01 & 43961 &  4999 & thin  & thick \\
 19.05 & 04:28 & 22:17 & 64109 & 21744 & thin  & thick \\ 
\noalign{\smallskip}
\hline
\end{tabular}
\end{center}
\label{tab:obs}
\end{table}

\section{pn-CCD and MOS-CCD images}
\label{sec:ima}

\begin{figure*}
\parbox{8.3cm}{\resizebox{\hsize}{!}{\includegraphics{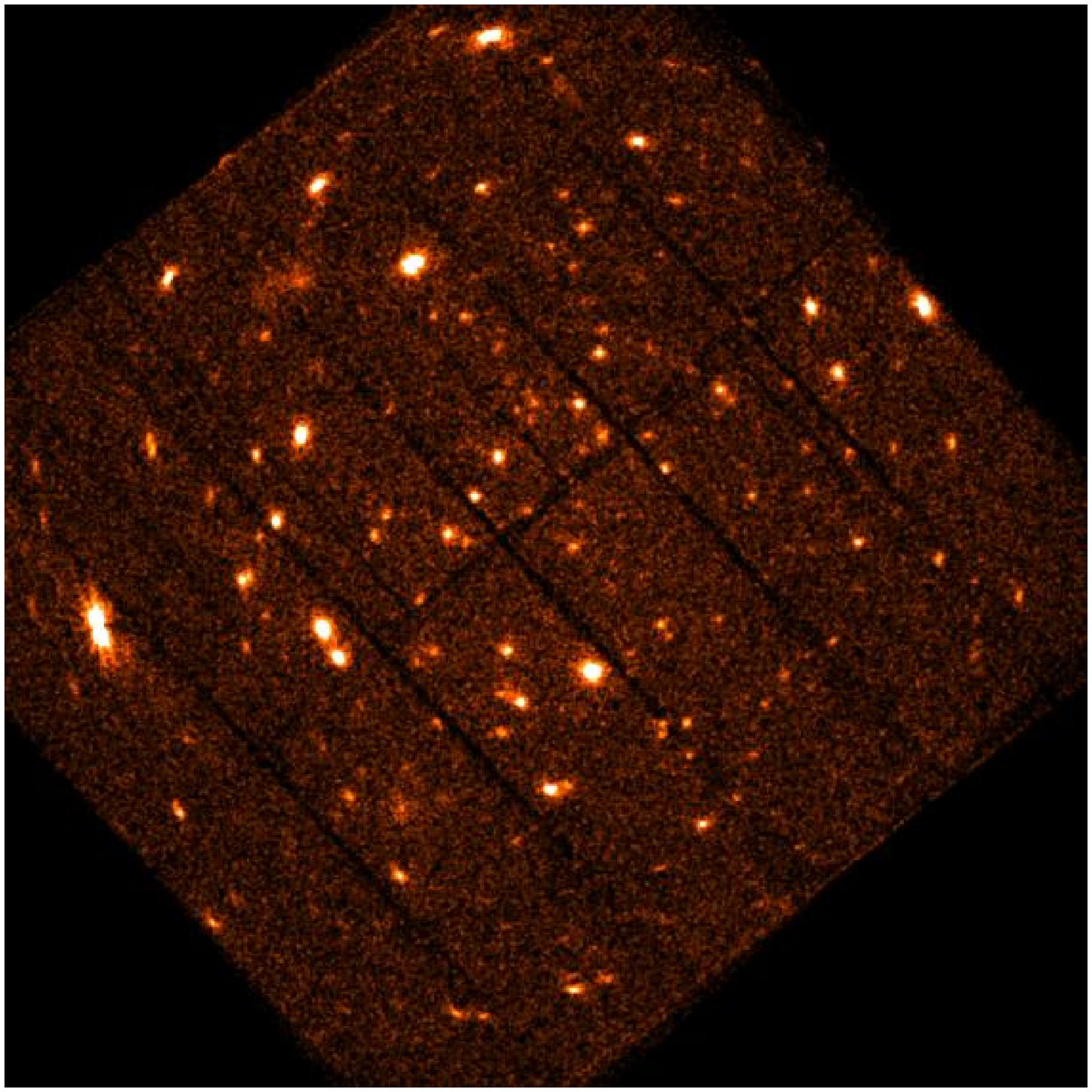}}}
\hfill
\parbox{8.3cm}{\resizebox{\hsize}{!}{\includegraphics{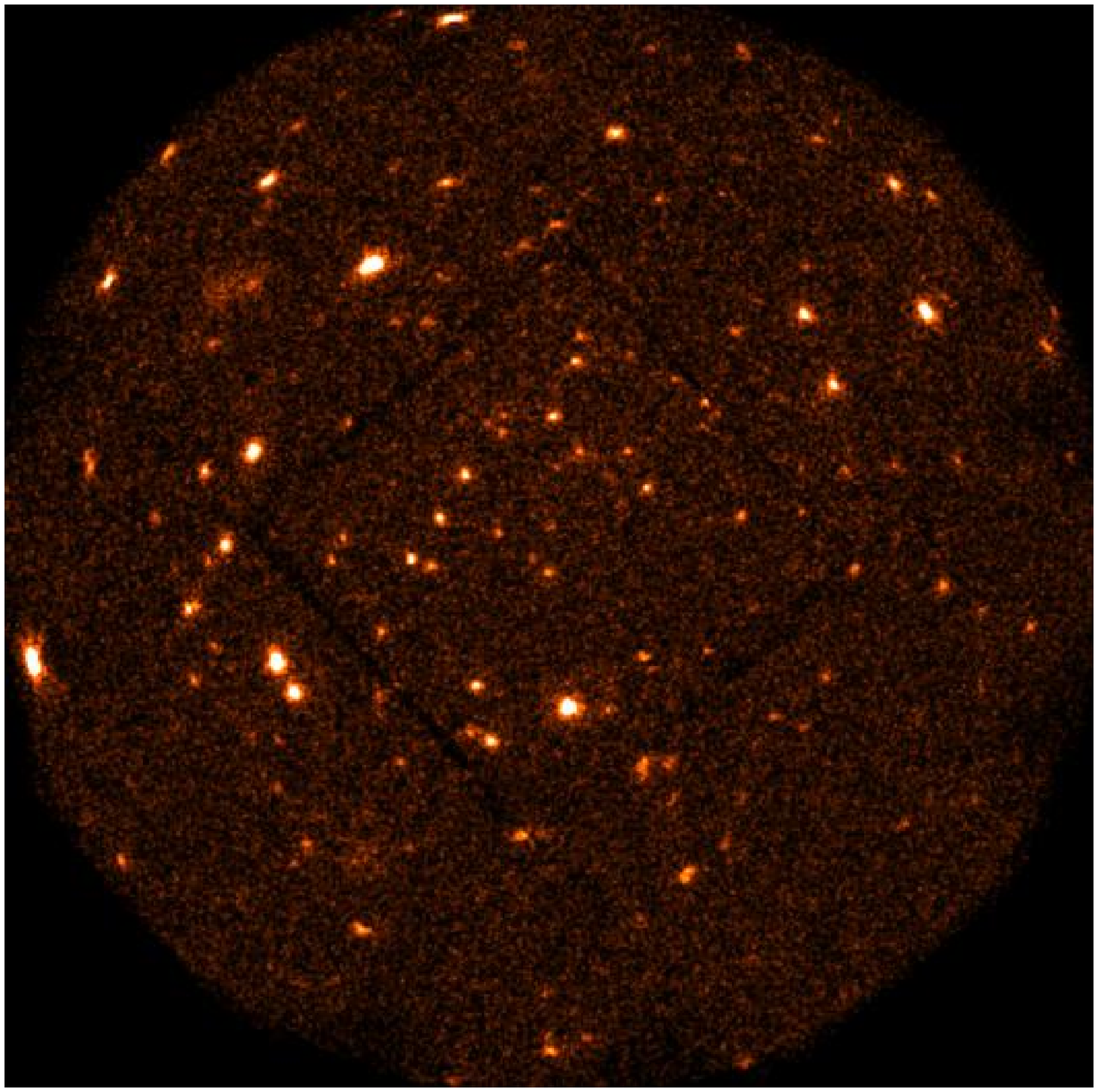}}}
\caption[]{X-ray images of the Lockman Hole obtained with the 
pn-CCD (left) and MOS 1+2 (right) cameras. The images have been
summed up over all {\em XMM}-revolutions  
for a total exposure of about 100 ksec and are accumulated in the 0.5-7 keV 
band. Both images are 30 $\times$ 30 arcmin across.
North is up and East is left.} 
\label{fig:raw}
\end{figure*}

\begin{figure*}
\parbox{8.3cm}{\resizebox{\hsize}{!}{\includegraphics{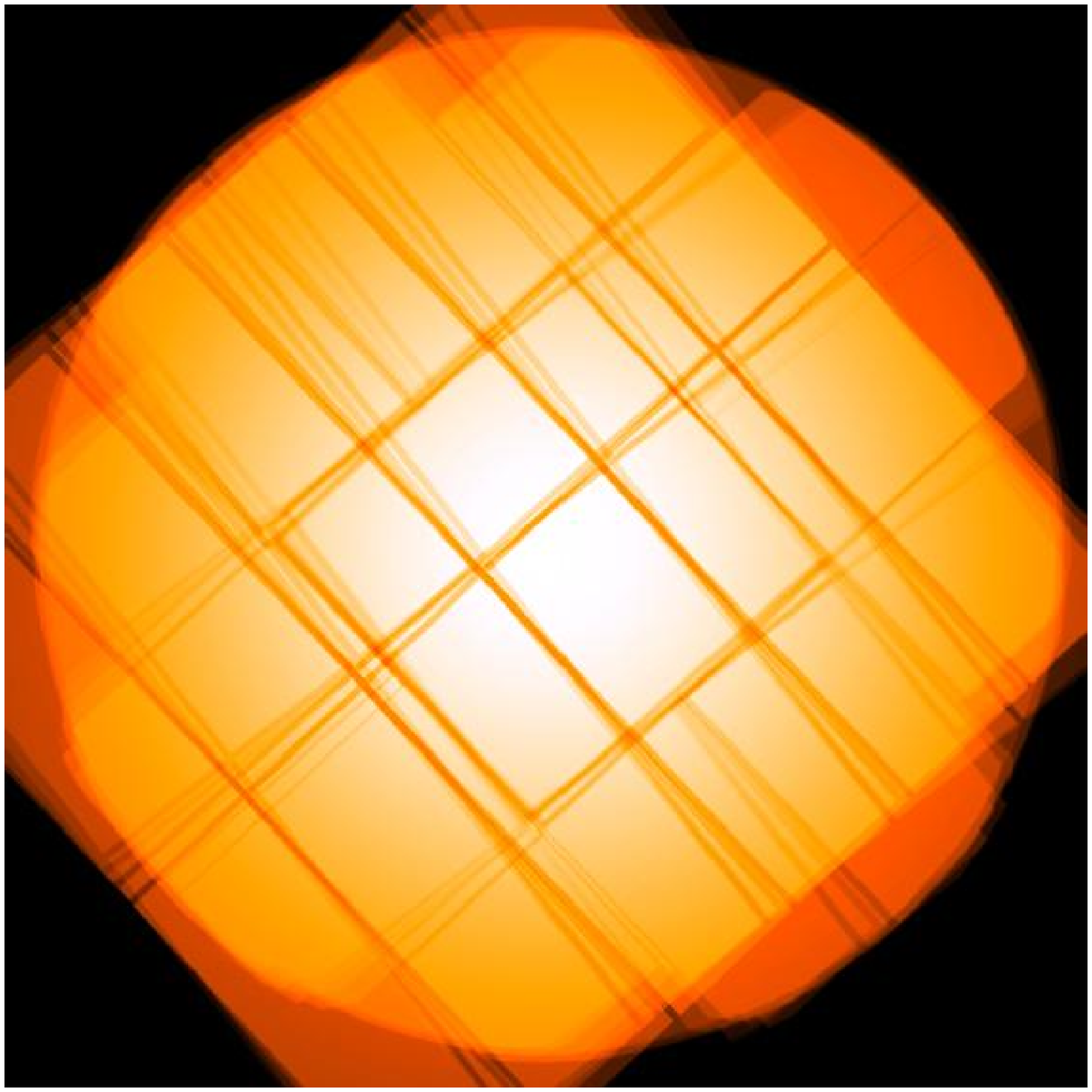}}}
\hfill
\parbox{8.3cm}{\resizebox{\hsize}{!}{\includegraphics{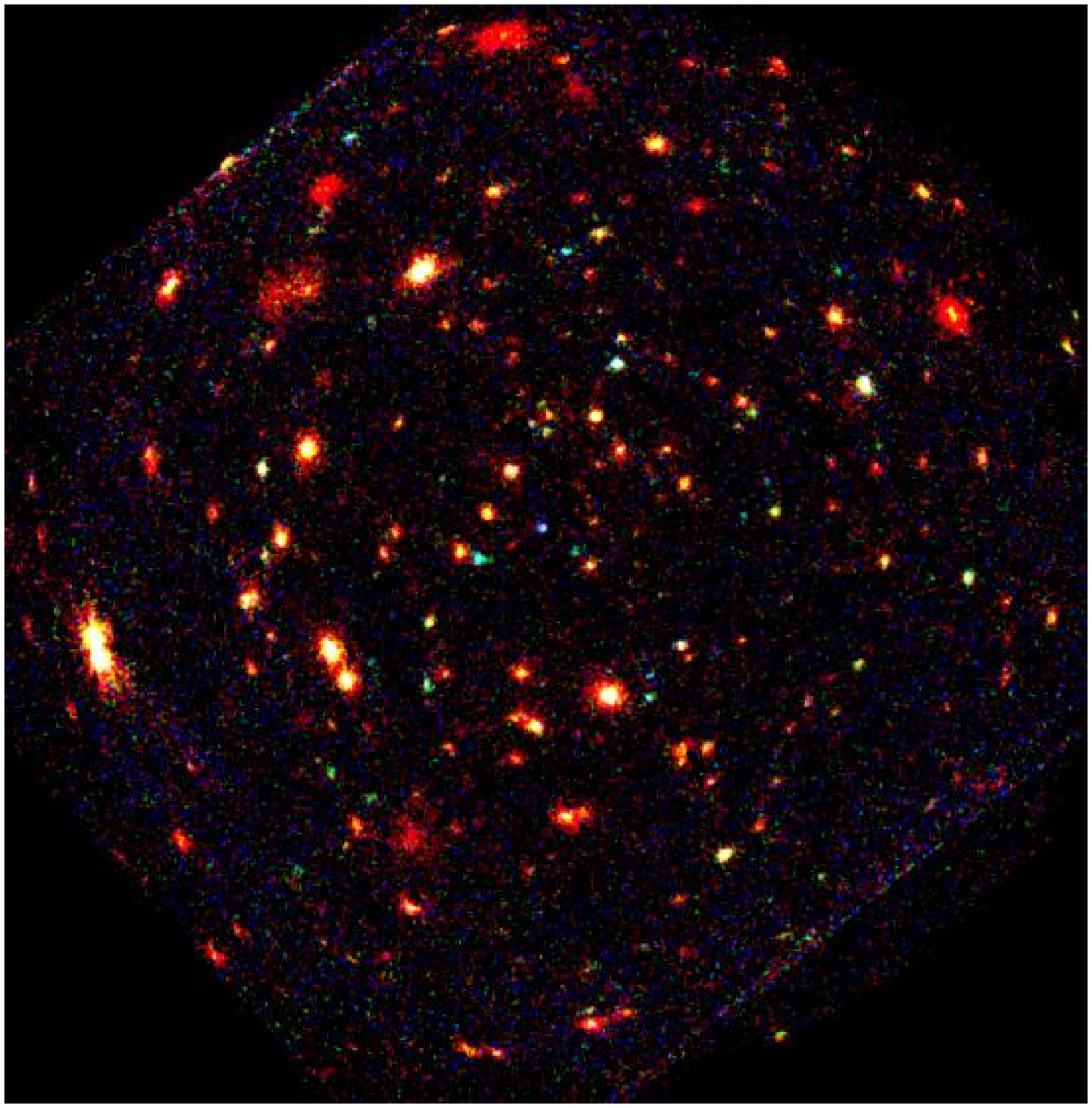}}}
\caption[]{a: (left) Combined exposure map for the PN and MOS images 
in Fig. \ref{fig:raw}. b: X-ray ``real-colour'' image of the 
combined and exposure corrected PN and MOS images. The colour 
refers to different X-ray energy bands: red, green and blue correspond to
the 0.5-2, 2-4.5 and 4.5-10 keV range, respectively.} 
\label{fig:exprgb}
\end{figure*}

Images in celestial coordinates with a pixel size of 2 arcsec have been 
accumulated in the 0.5-7 keV band for all three detectors. Fig. \ref{fig:raw}
shows a 
comparison of the image from the PN (left) and the MOS1+MOS2 camera (right). 
Albeit the variation in pointing direction between different revolutions, the 
shadows of the inter-CCD gaps appear in the images 
due to an interference between the variations in pointing direction and roll 
angles. An exposure map was calculated for the combination of pn-CCD plus 
MOS1 and MOS2 cameras (Fig. \ref{fig:exprgb}a). 

The brighter X-ray sources in the images have been already optically identified
from the {\em ROSAT} data (Schmidt et al. \cite{schmidt98}, Lehmann et al. 
\cite{
lehmann00,lehmann01}). X-ray sources identified with point-like optical objects 
(i.e. AGN or stars) have been used to check and correct the astrometry of the 
images. Positions of the brighter X-ray sources were centroided by an 
elliptical 2D-Gaussian fit. The FWHM of the point spread function in the 
center of the field was found to be 8.9 and 7.7 arcsec for the PN and MOS
images, respectively.
The X-ray source centroid sky coordinates calculated from the WCS keywords
were offset by 5$\arcsec$-25$\arcsec$ from the known optical counterparts. Thus a transversal
shift and a rotation angle were fit for each dataset. For the PN, satisfactory
fits were achieved by fixing the scale factor of the X-ray image to 1.0,
leading to residual systematic errors to $\sim1\arcsec$. For the MOS's,
introducing different scaling factors along the X- and Y-axes, in addition
to the transversal shift and rotation, improved the fits significantly. 
This reflects the fact that the geometrical layout of the
CCD chips in the MOS cameras had not yet been satisfactorily
established. 
The Lockman Hole observations can be used to improve the knowledge of the
layout, which then will be incorporated into future pipeline processings.
For our current analysis, we use these ad-hoc corrections,
which lead to residual systematic position errors of 1-3$\arcsec$.

\begin{figure*}
\parbox{5.9cm}{\resizebox{\hsize}{!}{\includegraphics{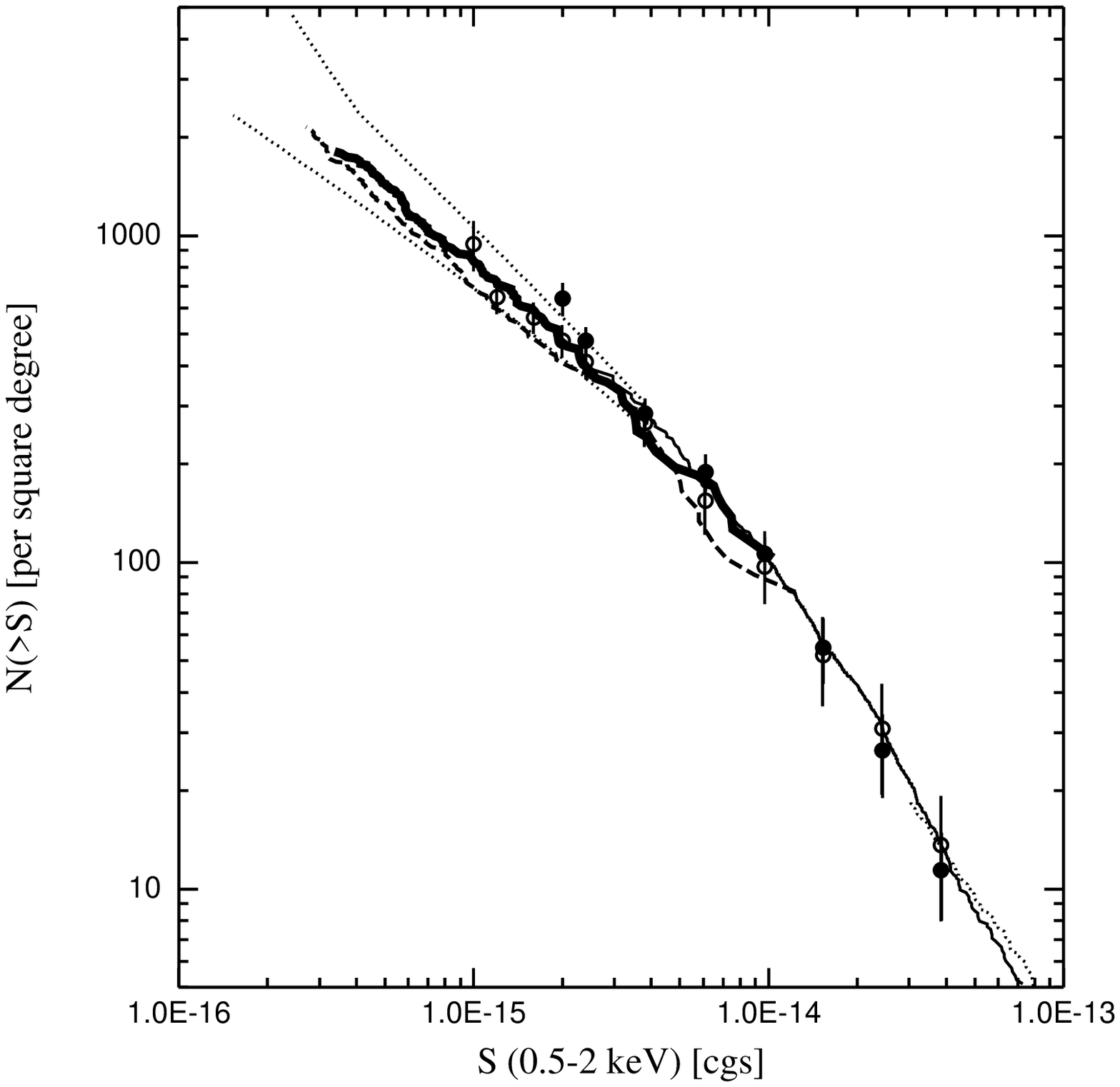}}}
\hfill
\parbox{5.9cm}{\resizebox{\hsize}{!}{\includegraphics{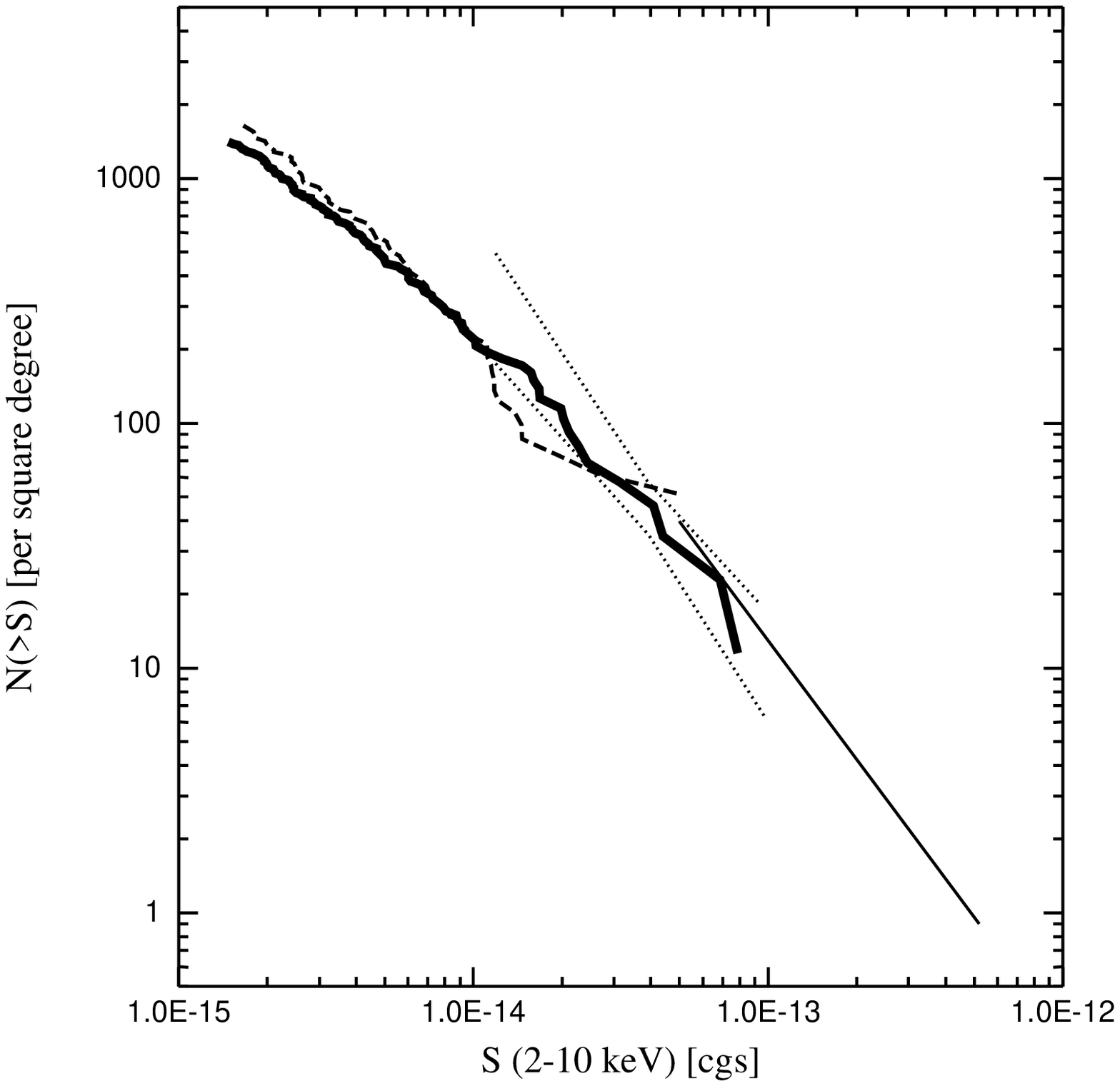}}}
\hfill
\parbox{5.9cm}{\resizebox{\hsize}{!}{\includegraphics{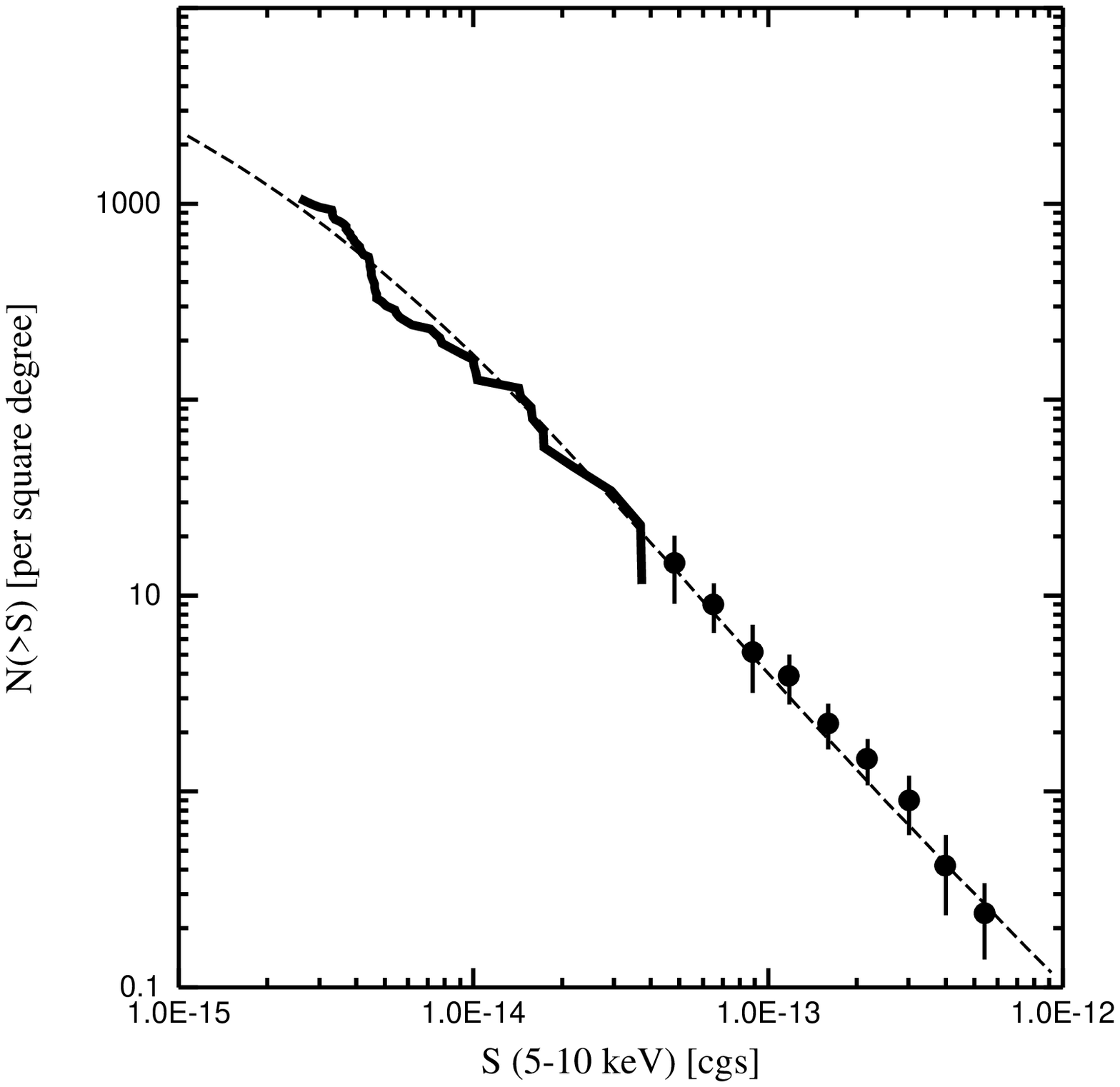}}}
\caption[]{Cumulative source counts N($\>$S) in the 0.5-2 keV (left), 2-10 keV
(middle) and 5-10 keV band (right). The {\em XMM} data are shown as thick solid
line. In the 0.5-2 keV band the data is compared with
the {\em ROSAT} source counts and fluctuation analysis (Hasinger et al. 
\cite{hasi98a})
and with the {\em Chandra} source counts (dashed line) of Giacconi et al. 
(\cite{giacconi00}). In the 
2-10 keV band the solid line at bright fluxes refers to the ASCA counts by
Cagnoni et al. (\cite{cagnoni98}), while the dashed line at faint fluxes again refer 
to the 
{\em Chandra} counts and the dotted region refers to the {\em BeppoSAX}
 fluctuation 
analysis by Perri \& Giommi (\cite{perri00}).  
In the 5-10 keV band the {\em XMM} counts are compared to the 
{\em BeppoSAX} log(N)-log(S) by Fiore et al. (\cite{fiore99}) and to a 
prediction based on the most recent background synthesis model 
by Gilli et al. (\cite{gilli99}).} 
\label{fig:logns}
\end{figure*}

\section{log(N)-log(S)}
\label{sec:logns}

As the residual systematic errors after the application of these 
astrometric corrections are below the width of the point-spread function, 
we could co-add the images of the different cameras.
Combined PN+MOS1+MOS2 
images were accumulated in the bands 0.2-0.5 keV, 0.5-2 keV, 2-4.5 keV and 
4.5-10 keV, respectively. Fig. \ref{fig:exprgb}b shows the exposure-corrected
image of all 
cameras combined in an X-ray ``real-colour'' representation. The red, green 
and blue colours refer to the 
0.5-2, 2-4.5 and 4.5-10 keV images, respectively. A population of green and 
blue objects is showing up in this image, i.e. obscured faint X-ray sources 
which have been postulated by the X-ray background population synthesis models.
There are several diffuse sources with red colours which 
are X-ray clusters of galaxies already identified from the {\em ROSAT} data
(Schmidt et al. \cite{schmidt98}, Hasinger et al.
\cite{hasi98c}, Thompson et al. \cite{thompson00}).

The SAS source detection algorithms
have been applied to the data. These are an improved variant of the 
{\em ROSAT}
source detection algorithms described in Hasinger et al. (\cite{hasi98a})
consisting of simple sliding window box detection algorithms, using either a 
local background estimate (LDETECT) or a background map derived from 
the images smoothed by a bi-cubic spline function after bright sources have 
been removed (MDETECT) as well as a multi-ML source detection and parameter 
estimation task. The main improvement for the {\em XMM} data is, that these 
algorithms are run simultaneously for several independent energy bands, 
keeping the source positions and extent fixed for all energy bands, while 
adding the 
source existence likelihoods from the individual energy bands together.
This improved algorithm is both more sensitive and less subject to source 
confusion and yields source count rates in all energy bands. 

Source detections were accepted with likelihood values above 10 (about 
4$\sigma$) and inside an off-axis angle of 10 arcmin. 
The resulting detection statistics are given in Tab. \ref{tab:det}.
The raw source count 
rates have been converted to X-ray fluxes by applying a correction for 
vignetting (up to a factor of 2 at an off-axis angle of 10$\arcmin$) and dead time 
plus out of time events 
(assumed to be 8\% for the combination of all three detectors) 
as well as the counts-to-flux conversion factor (ECF) according to Tab.
\ref{tab:det}. The average energy conversion factors for the whole 
observation have been computed using the most recent response matrices
weighted with the respective exposure times for the three different detectors.
The vignetting function has been assumed to depend linearly on
off-axis angle, while the azimuthal dependence expected for the MOS
cameras has been ignored. 
Due to the still preliminary status of the current EPIC calibration and the 
simplifying assumptions made here we have to assume systematic 
flux errors on the order of 10\% in addition to the uncertainty
due to the range in possible photon indices (see Tab. \ref{tab:det}).

\begin{table}
\caption[]{Detection results for different energy bands}
\begin{center}
\begin{tabular}{lrrlrr}
\hline
\noalign{\smallskip}
Band$^a$  & $\Gamma$$^b$ & $ECF^c$   & $S_{lim}$$^d$ & $N_{src}$$^e$ & $N(>S)$$^f$\\
\noalign{\smallskip}
\hline
\noalign{\smallskip}
0.2-0.5 & 2.0$\pm$0.5 &  7.16$\pm$1.01 & 0.40 & 120 & 1380 \\ 
0.5-2   & 2.0$\pm$0.5 & 10.20$\pm$0.04 & 0.31 & 148 & 1800 \\
2-10    & 2.0$\pm$0.5 &  1.79$\pm$0.40 & 1.4  & 112 & 1400 \\ 
5-10    & 1.6$\pm$0.5 &  1.28$\pm$0.11 & 2.4  &  61 & 1060 \\
\noalign{\smallskip}
\hline
\end{tabular}
\end{center}
$^a$ energy band in keV in which the flux is given (see text)  \\
$^b$ assumed range in photon index\\
$^c$ energy conversion factor in cts~s$^{-1}$ per $10^{-11}~{\rm erg}~{\rm cm}^{-2}~{\rm s}^{-1}$\\
$^d$ minimum detected flux in $10^{-15}~{\rm erg}~{\rm cm}^{-2}~{\rm s}^{-1}$\\
$^e$ number of sources detected within 10 arcmin radius\\
$^f$ source density in deg$^{-2}$\\
\label{tab:det}
\end{table}

The  corresponding cumulative log(N)-log(S) distributions are shown in 
Fig. \ref{fig:logns}. 
In the soft band (0.5-2 keV) the data reach a flux limit about a 
factor of three deeper than the {\em ROSAT} HRI survey (Hasinger et al. 
\cite{hasi98a}), 
and are about 50\% less sensitive than the recent {\em Chandra} surveys 
(Mushotzky et al. \cite{mushotzky00}; Giacconi et al. \cite{giacconi00}). 
In the 2-10 keV band the {\em XMM} data are
as deep as the published {\em Chandra} surveys. 
In the very
hard 5-10 keV band, which has been pioneered by the {\em BeppoSAX}
observations (Fiore et al. \cite{fiore99}), 
{\em XMM} is entering new territory, reaching more than a magnitude
deeper than {\em BeppoSAX}. 
As described above, the flux conversion still has to be regarded as uncertain 
by at least 10\%. Also no corrections for confusion or Eddington biases have 
been made 
yet (see e.g. discussion in Hasinger et al. \cite{hasi98a}). Nevertheless,  
the comparison of the {\em XMM}
source counts with the {\em Chandra} and {\em ROSAT} data, in particular in 
the 0.5-2 keV band, 
indicates that confusion is not a severe problem for the source counts.

\begin{figure*}
\parbox{8.0cm}{\resizebox{\hsize}{!}{\includegraphics{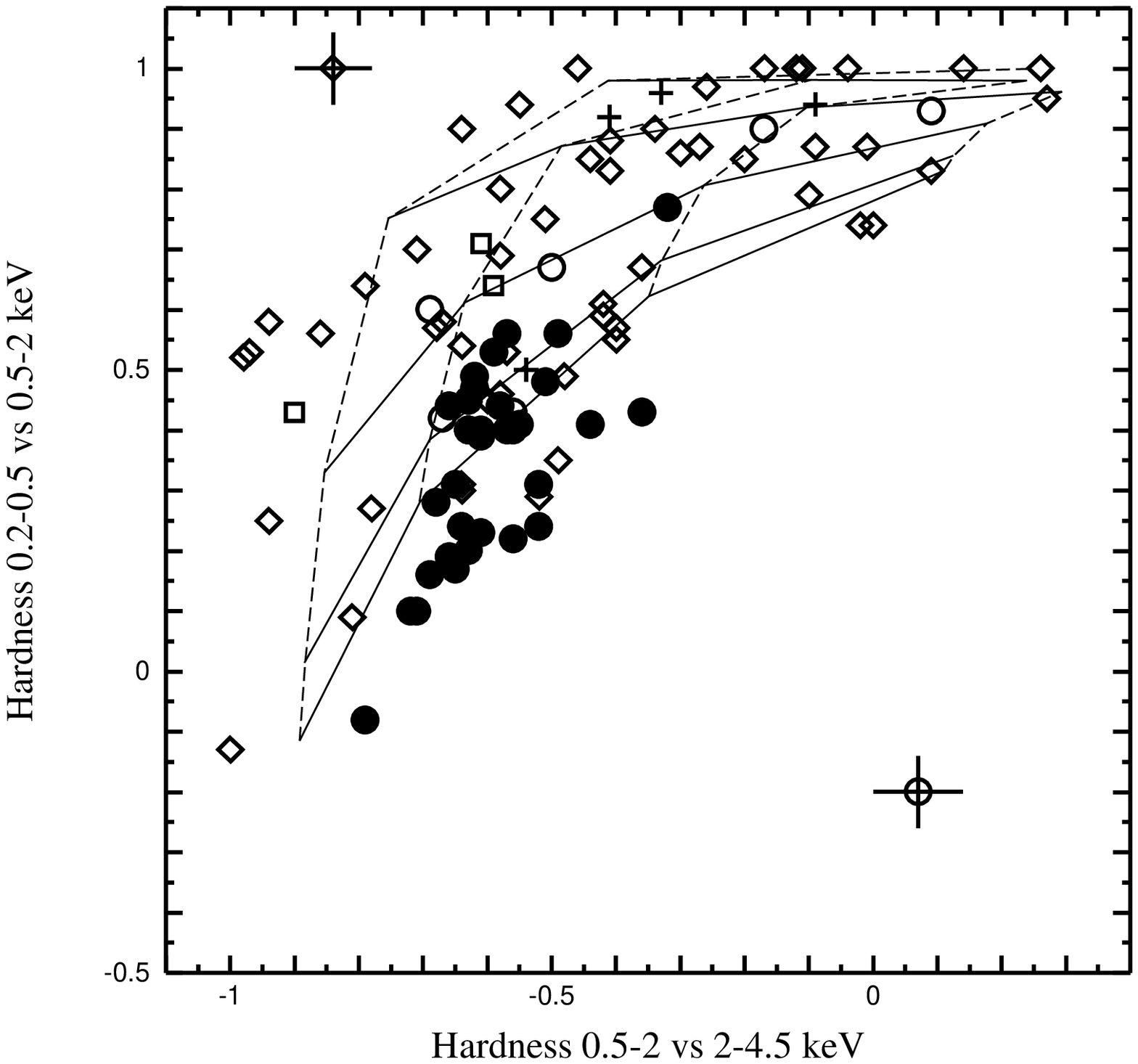}}}
\hfill
\parbox{8.0cm}{\resizebox{\hsize}{!}{\includegraphics{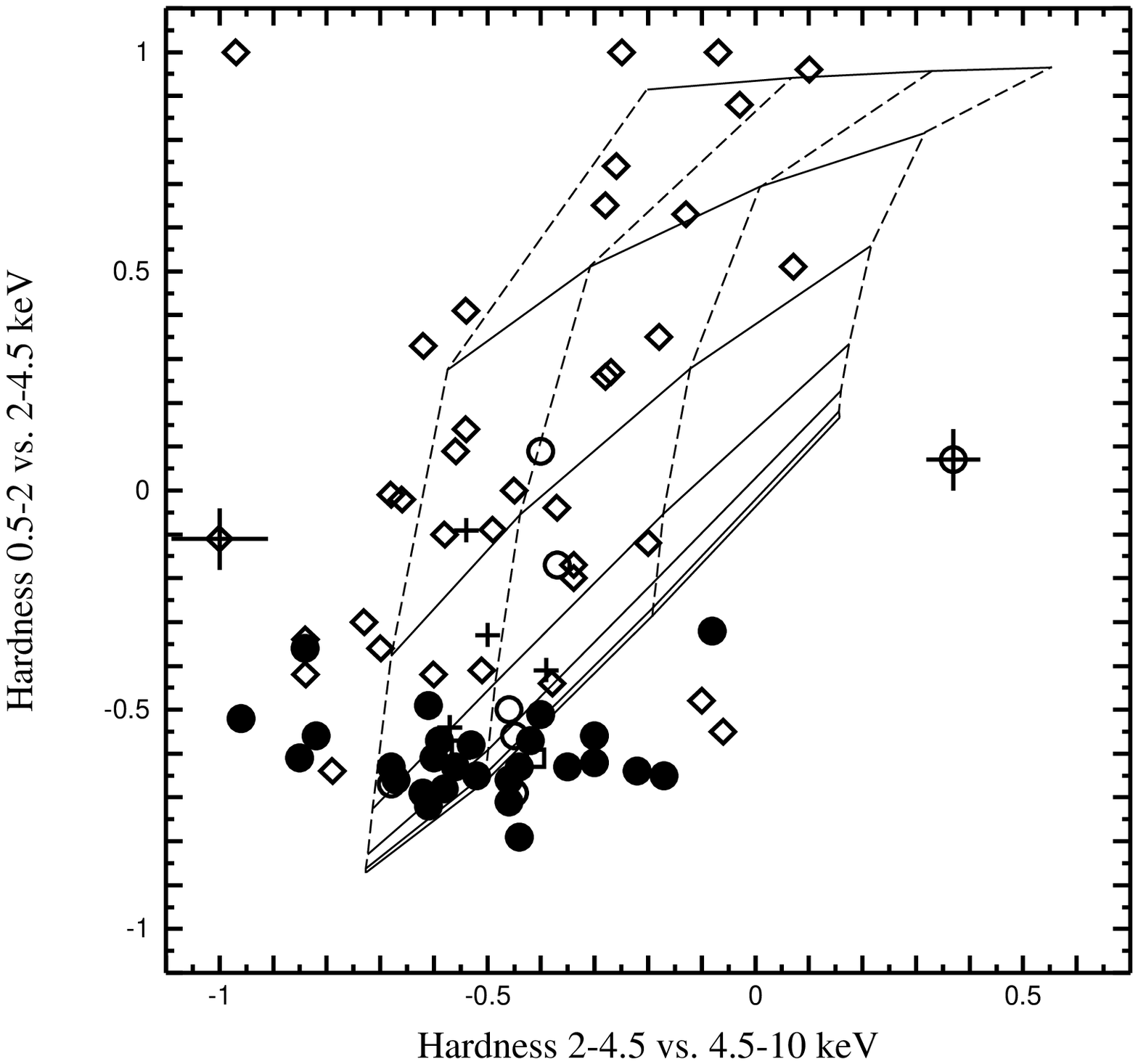}}}
\caption[]{X-ray spectral diagnostic diagrams based on hardness ratios
(see text). The symbols refer to different classes of objects detected in 
the {\em XMM} deep survey of the Lockman Hole: filled circles are 
type-1 AGN spectroscopically identified in the {\em ROSAT} ultradeep HRI 
survey (Lehmann et al. \cite{lehmann01}), open circles are type-2 AGN,
correspondingly. Open squares are clusters or groups of galaxies and
plus signs are spectroscopically not identified {\em ROSAT} sources 
(most likely type 2 AGN, see Lehmann et al., \cite{lehmann00}) with
photometric redshifts. Open diamonds refer to the newly detected 
{\em XMM} sources. Two representative error bars are shown; 
for clarity only sources with hardness errors less than 0.1 are
plotted. The grid gives the expected hardness ratios for 
power law models with photon indices $\Gamma =$ 0, 1, 2 and 3 
(dashed lines) and neutral 
hydrogen absorption (in the observed frame) of $log N_H $ between 
20 and 23 in steps of 0.5 (solid lines). 
} 
\label{fig:col}
\end{figure*}

\section{X-ray colour-colour diagrams}
\label{sec:colcol}

The unprecedented combination of high throughput and wide energy 
band makes {\em XMM} uniquely suited to classify sources based on their
X-ray spectra alone. The large number of X-ray sources in
the Lockman Hole which already have spectroscopic identifications and redshifts 
based on the {\em ROSAT} surveys can be used as a training set for the 
classification of the still unidentified {\em XMM} sources in the Lockman 
Hole, but also in other XMM fields.

From the count rates in the four independent energy bands used for source 
detection (see Sect. \ref{sec:logns}) we have calculated three statistically
independent hardness ratios, HR1, HR2 and HR3, according to the formula 
$HR = {(H-S)/(H+S)}$, where $H$ and $S$ correspond to the counts in the
harder and softer energy bands, respectively (see Tab. \ref{tab:det}). Fig. 
\ref{fig:col} shows two X-ray colour-colour diagrams with different 
symbols for various source classes superposed on a model grid for 
absorbed power law spectra.    
The type-1 AGN known from {\em ROSAT} populate a narrow, relatively
soft range in both diagrams, consistent with a photon index of 
$\Gamma \approx 2$ and typically low apparent absorption column densities 
($\log(N_H)<21.5~{\rm cm}^{-2}$).  The new {\em XMM} sources, together 
with a number of previously known type-2 {\em ROSAT} AGN scatter 
over a much wider area. Although the new {\em XMM} sources are typically
fainter than the {\em ROSAT} sources and therefore have larger hardness ratio
errors, the new population is considerably harder in both diagrams.  
A comparison of the source colours with the underlying model grids 
shows, that the hardening is mainly due to apparent absorption column 
densities of $\log(N_H)>21.5~{\rm cm}^{-2}$ on top of relatively soft 
spectra and not due to intrinsically hard power law indices. 
Some sources fall outside the model grid. In one case, a ROSAT Seyfert-2 
galaxy which shows the largest HR3, this is probably due to a very soft 
component on top of a heavily absorbed power law. In some cases the 
hardness ratios might be affected by source confusion.

\section{Discussion and Conclusions}
\label{sec:disc}

We have shown the first log(N)-log(S) relations based on the 
{\em XMM-Newton} observatory. Given the still existing systematic 
uncertainties, the data is fully consistent with the {\em 
ROSAT} and {\em Chandra} source counts in the 0.5-2 keV band. 
This demonstrates on one hand that the combination of EPIC detectors is
not yet confusion limited in a 100 ksec observation, on the other 
hand that cosmic variance between different fields does 
not affect the source counts significantly at the currently achieved flux
levels, at least not in the soft band.  In the 2-10 keV band there is an 
inconsistency of about 40\% 
between the two recent {\em Chandra} datasets by Mushotzky et al. 
\cite{mushotzky00} and Giacconi et al. \cite{giacconi00}, the 
latter one having a lower normalisation. The new {\em XMM} data are 
consistent with the Giacconi et al. log(N)-log(S), maybe even somewhat
flatter, and clearly confirm a break in the
slope compared to the quasi-Euclidean behaviour at brighter fluxes. 
In the 5-10 keV band the {\em XMM} data go more than an order of 
magnitude deeper than the previous {\em BeppoSAX} counts
(Fiore et al. \cite{fiore99}). There is so far relatively 
little deviation from a Euclidean slope and the data is fully consistent with 
the predictions from recent population synthesis models for the X-ray 
background (Gilli et al. \cite{gilli99}). 
Adding up the source counts, we resolve about 60\% of the 5-10 keV X-ray
background.

The diagnostic power of {\em XMM} lies in its wide energy band and its
unprecedented sensitivity in the hard band. With the help of X-ray
colour-colour diagrams and the ``training set'' of about 60 previously
identified {\em ROSAT} sources in the same field it is possible to
characterise the 
new XMM sources as typically harder, probably intrinsically absorbed
sources. A small number of objects with similar X-ray colours has already 
been identified in the deepest ROSAT survey (Lehmann et al. \cite{lehmann00}), 
they are type-2 Seyferts or unidentified objects with extremely red 
optical/NIR colours ($R-K > 5$). The new XMM source population is 
therefore very likely dominated by obscured AGN, as predicted by the 
AGN population synthesis models for the X-ray background.

\begin{acknowledgements}
      We thank the whole {\em XMM-Newton} team for providing such a wonderful 
      observatory and producing well-calibrated data at an early time in 
      the mission. 
      We thank R. Gilli for providing the model prediction in Fig. 
      \ref{fig:logns}. We acknowledge very useful coments by the referee, 
      G. Zamorani. Part of this work was supported by the German
      \emph{Deut\-sches Zentrum f\"ur Luft- und Raumfahrt, DLR\/} project
      numbers 50 OX 9801 and 50 OR 9908.
\end{acknowledgements}

\end{document}